\documentstyle[12pt,epsf,titlepage,fleqn]{article}

\newcommand{\be}{\begin{equation}}
\newcommand{\ee}{\end{equation}}
\newcommand{\ba}{\begin{eqnarray}}
\newcommand{\ea}{\end{eqnarray}}
\newcommand{\om}{\omega}
\newcommand{\q}{\hat{q}}
\newcommand{\p}{\hat{\pi}}

\newcommand{\oh}{\frac{1}{2}}

\begin{document}

\pagebreak
\begin{titlepage}
\vspace*{0.1truecm}
\begin{center}
\vskip 1.0cm
{\large\bf  BROWNIAN MOTION OF SOLITONS IN THE $\Phi^4$ MODEL}

\vskip 2.0cm

{\large F. ALDABE\footnote{E-mail: aldabe@ictp.trieste.it}}

{\large\em International Center for Theoretical Physics\\
P.O. Box 586, 34100 Trieste, Italy}\\
\vspace{.5in}
\today \\
\vspace{.5in}
{\bf ABSTRACT}\\
\begin{quotation}
\noindent
We derive an expression for the correlation function of the random force on
a soliton which is consistent with the constraints needed to integrate out
the zero modes which appear due to the broken translational symmetry
of the soliton solution.  It is shown that when the constraint does not commute
with
the operator which defines the correlation function, i.e. when the operator
is not  physical,
 only low frequency phonons
contributions may be considered.  On the contrary,
when the
correlation function of the random force on the soliton is constructed
with physical operators one may also
include in a correct manner the contributions from the optical phonons.

\end{quotation}

\end{center}
\vfill
\end{titlepage}

\section{Introduction}

Although the $\phi^4$ is not an adequate model
to describe solitons and their motion in polyacetylene because it does not
include
retardation effects, it is a model simple enough to make possible the study of
soliton
diffusion \cite{Wad}.
In this model, the soliton is taken as the classical solution of a
non linear differential equation.  This solution is stable because it is
possible to define a topological charge which commutes to leading order
with the Hamiltonian describing the soliton.  It is then impossible to
deform the soliton solution to a trivial solution in a continuous manner.
The quantum fluctuations about the soliton are then the phonons of the
model.  Since the Hamiltonian is non linear, there are interactions between
the soliton and the phonons.

One may then consider how these phonons give rise to a Brownian motion of the
soliton.
The observable which measures how
the Brownian motion takes place is the diffusion operator.  The diffusion
operator has a diffusion constant which is proportional to the square
of the phonon number.  Thus, at zero temperature we expect the soliton
to undergo a mild Brownian motion due to quantum effects.  As temperature
increases, soft frequency phonons will contribute to the diffusion of the
soliton.  As temperature increases even more, optical phonons will come
into the picture and contribute to the Brownian motion of the soliton.
Thus, in order to describe soliton diffusion at high temperature in
a correct manner it is necessary to take into account the contribution
of optical phonons.  Experimental values of the extension of the soliton
are of the order of the wavelength of optical phonons.  It is
thus important that the information of the profile of the soliton
entering the correlation function for the random force be in
agreement with that of the classical solution.

The problem of soliton diffusion has been studied in \cite{Wada}, were used was
made of the formalism presented in \cite{Mori} to calculate the contribution to
the diffusion of the soliton due to phonons.
However, the analysis done in \cite{Wada} does not take into
account in a proper manner the shape of the soliton.  Thus, the contribution
from the optical phonons to the correlation function used in \cite{Wada}
is incorrect.  This follows from the  analysis done below were we show that
the correlation function as defined by Mori \cite{Mori} is  constructed from
an unphysical
operator.  This operator is not physical in soliton models
 because these systems have zero modes associated to the loss of translational
invariance of the classical solution.  In order to integrate out this zero
mode and obtain a theory free of infrared singularities
 we must introduce a collective
coordinate.  In doing so we enlarge the phase space, and to recover
the physical theory we must make use of a constraint which appear
in the definition of the collective momentum.  That is, the collective
momentum is not the time derivative of the collective coordinate.  Rather,
it is a function of the fluctuations and classical solution.  This relation
imposes
a
constraint.  And this constraint does not commute with the collective
coordinate.
Thus, the collective coordinate is not physical.
 This can alternatively be understood by looking at
the original phase space.  This space is embedded in a large space which
also contains the collective coordinate which was not present in the
original theory.  Despite subtleties which involve certain transformations
among the variables in the enlarged phase space, the collective coordinate
is an artificial operator which is not physical.

The reason is that one
cannot simply measure the position of the soliton.  The best one
can do is to measure the energy density of the soliton.  Thus, one can
effectively measure the classical soliton solution.  Where the solution
vanishes we define the position of the soliton.  In fact, if one
expands the soliton solution in a Taylor series in the collective
coordinate, she or he will find that the leading term is nothing
but the collective coordinate.  However, the soliton solution is not physical
either.  To be physical one must add the fluctuations about the soliton.
Thus, one must replace the collective coordinates appearing in the correlation
function of the random forces which act on the soliton  defined in \cite{Mori}
by the field appearing in the theory.

When such a replacement is made, we find that the information entering
the correlation function of the random forces acting on the soliton will
contain all the information of the soliton profile and not just the
first term in the Taylor series in the collective coordinate.
Retaining only the first term is equivalent to assuming that the energy
density of the soliton is evenly distributed throughout the polymer.  Thus,
low frequency phonons will not see the shape  of the soliton
but neither will the optical
phonons.  While the soft phonon have wavelength much larger than the
measured extension of the soliton and thus do not effectively
see the soliton, this is not the case for optical phonons whose wavelength
is of the order of the extension of the energy density of the soliton.  Thus by
keeping only the first term in the Taylor series, e.i.
the collective coordinate,
the contribution of the optical phonons to the correlation function of the
random forces acting on the soliton is incorrect.  Only when we retain the
remaining terms in the expansion will the contribution of these phonons
be correct.  Thus, conclusions regarding soliton diffusion at high
temperatures using the diffusion equation in \cite{Wada} must be
reconsidered and effects from optical phonons recalculated.

\section{The Model}
We consider the propagation of a soliton in a box of length $L$.
  The Lagrangian is
\be
L(\Phi,\Psi)=L(\Phi)+L(\Psi)+L_{int}
\ee
and it is invariant under translations.  The first, second and third
terms on the r.h.s refer to the bosonic term, fermionic term
 and interaction term between them,
respectively.
The fermions will not play a role in our discussion.
We thus concentrate on the bosonic sector.  The Lagrangian for this
sector reads
\be
L=\int  dx
(\dot{\Phi}^2-\Phi'^2+\Phi^2-\frac{1}{2\lambda^2}\Phi^4-\frac{\lambda^2}{2})
\ee

The classical equation of motion for this system is
\be
\ddot{\phi}-\phi''+V_1(\phi)=0.
\ee
where $V_n=\partial^n V/\partial\phi^n$.   Static solutions to this equations
can be classified
according their translational invariance.  For example, the constant solution
\be
\phi_{co}=\lambda
\ee
has translational invariance, and vanishing topological charge.  The Solution
without translational
invariance
\be
\phi_c=\lambda \ tanh(\frac{x}{\sqrt{2}}),\label{ex}
\ee
satisfies
\be
\phi_c(x)\to{\phi}_c (x+X)\ne\phi(x).
\ee
The solution $\phi_c$
is stable since it has topological charge equal to one and cannot decay without
changing this
value.

The loss of translational invariance of the classical solutions lead to the
appearance of
zero modes in the spectrum of the fluctuations.  The spectrum of the
fluctuations $\q$ with
\be
\Phi=\phi_c+\q.
\ee
is obtained from the linearized equation of motion for the fluctuations
\be
-\ddot{\q}+\q''+V_2|_{\phi_c}\q^2=0.\label{eomf}
\ee
Expanding the fluctuations in normal modes
\be
\q=\sum_n\frac{i}{\sqrt{2\om_n}}(\psi_n a^+_n+h.c.),
\ee
we see that among the solutions $\psi_n$ to (\ref{eomf}) there is a normalized
solution
\be
\psi_1=\frac{\phi'_c}{\sqrt{M}}
\ee
where $M$ is the mass of the soliton.  The eigenfrequency of this solution is
$\om_1=0$.
This is the zero mode.  In doing a perturbative treatment of an observable
we will
encounter infrared divergences coming from the contributions of this zero mode.
In order to integrate out these zero modes
and obtain an infrared divergent free theory we must make use of collective
coordinates.

The solutions $\psi_n$ to the equation (\ref{eomf}) satisfy
\ba
\delta(x-y)&=&\sum_n \psi_n(x)\psi_n^*(y)\\
\int  \psi_n\psi^*_m&=&\delta_{nm}\label{ort}
\ea
With the help of these relations we can write the quadratic Hamiltonian in the
form
\be
H=\sum_n\om_n(a^+_na_n+\oh)+(p^{(1)})^2
\ee
where the summation in the first term excludes the zero mode and the
last term depends only on the zero mode creation and annihilation operators
$a^+_1$ and $a_1$
\be
p^{(1)}=\psi_1\sqrt{\frac{\om_1}{2}}(a_1+a^+_1)
\ee

\section{The Collective Coordinates}

We gauge the translational invariance \cite{bk}.
We do this by raising the parameter $X$ to a dynamical variable $X(t)$ which we
will
refer to as collective coordinate.  The field will have a dependence on $X(t)$
\be
\Phi(x)\to\Phi(x+X(t))
\ee
Considering
this dependence in the Lagrangian we find that we recover the equation of
motion for
$\Phi$.  In going to the Hamiltonian formalism, the conjugate of $\Phi$ is
\be
\Pi:=\frac{\partial L}{\partial\dot{\Phi}}=\p
\ee
where ${[\p(x),\q(y)]}=-i\delta(x-y)$.
We also acquire an  additional equation
which follows from the definition of the collective momentum.
\be
 P:=
\frac{\partial L}{\partial\dot{X}}=\int \Phi'\p
\ee
This is a constraint which allows us to integrate the zero mode without
encountering
infrared divergences.  The constraint is best put in the form
\be
f=\int\phi_c' p+\q'\p- P
\ee
Standard treatment of systems with collective coordinates requires that
physical
operators commute with the constraint and that physical states be annihilated
by
the constraint.  The reason for this requirement is simple.  We have embedded
our
theory, which was defined in a phase space which contains $\Phi$ and its
conjugate only,
into a larger phase space which also includes the collective coordinate
and its
conjugate.  The embedding is performed by the gauge transformations whose
generator is $f$.  We require that physical operators, those which were defined
in the original theory, be independent of the collective coordinate before the
embedding.  This is equivalent to requiring that the operator commute with
the constraint after the operator has been embedded.

Examples of
operators which are physical are the Hamiltonian $H$ and the field $\Phi$
for which it holds
\be
{[H,f]}=0,\;\;\;{[\Phi,f]}=0.
\ee
Operators which are not physical are the fluctuations $\q$ and the collective
coordinate $X$ since the commutators
\be
[f,X]=-i\;\;\;\;[f,\q(x)]=-i\Phi'(x)
\ee
do not vanish.
To determine if $\dot{X}$ is a physical operator we first write the Hamiltonian
 after  integrating out the zero mode.
The infrared free collective Hamiltonian can be written as \cite{Gjs}
\be
H_{coll}=\oh \{( P-\int\q'\p)^2,\frac{1}{2M(1+\frac{\phi_c'\q'}{M})^2}\}_+.
\ee
Using the definition
\be
\dot{X}=\frac{\partial H}{\partial  P},
\ee
we obtain an expression for $\dot{X}$
\be
\dot{X}=
\oh\{( P-\int\q'\p),\frac{1}{M(1+\frac{\phi_c'\q'}{M})^2}\}_+.\label{e23}
\ee
It is a simple exercise to check that the commutator
\be
[f,\dot{X}]\ne0.
\ee
Thus, the operator $\dot{X}$ does not have a physical meaning.  A similar
exercise for $\ddot{X}$ shows that neither this operator is physical.

\section{The Diffusion Equation}
The Fourier transform of the correlation function of the random force
on the soliton
was written in \cite{Wada}  as
\be
\Gamma(\om)=
\int \frac{<\ddot {X}(t)\ddot{X}(0)>}{<\dot {X}(t)\dot{X}(0)>}e^{-i \om t} dt.
\label{dif}
\ee
Equation (\ref{dif}) is related to the diffusion equation for the soliton
\be
D(\om)=\frac{\alpha}{i\om +\Gamma(\om)}
.\label{def}
\ee
Where $\alpha$ is the thermal average of the square of the soliton velocity.
For adiabatic phonons, equation (\ref{dif}) is a good approximation.
The reason for this
follows from the fact that the operator $X$ and its time derivatives are not
physical operator.
Thus the equations (\ref{dif}) and (\ref{def}) do not have a physical meaning.

The natural question which arises is which is the minimal operator that must be
added to
the operator $X$ to have a physical operator.  To leading order, the constraint
implies
\be
 P=\int\phi_c'\p
\ee
Since $ P$ is the conjugate of $X$ it follows that
\be
X(t)=\frac{\int\phi_c' \q}{M}.\label{xq}
\ee
The operator in the r.h.s. of  (\ref{xq}) is unphysical and the minimal
operator which
includes the r.h.s of (\ref{xq}) and is physical at the same time  is
\be
\phi=\phi_c+\q_{re}+\psi_1 X(t)
=\phi_c+\q_{re}+\psi_1\int\psi_1\q=\phi_c+\q.\label{phi}
\ee
where in the second term use was made of (\ref{ort}) and
\be
q_{re}(x)=\sum_{n\ne 1}\psi_n(x)\int \psi^*_n q.
\ee
The equation (\ref{phi})
is a sensible choice because the equation (\ref{def}) refers to the diffusion
of
the soliton and not of the collective coordinate of the soliton.
 However, the collective coordinate is a good but
unphysical approximation to the soliton.  To see that this is an approximation
we write the equation (\ref{dif}) in terms of the soliton field
\be
\Gamma(\om)=\int \frac{<\ddot {\Phi}(t)\ddot{\Phi}(0)>}{<\dot
{\Phi}(t)\dot{\Phi}(0)>}e^{-i \om t} dt.
\label{diff}
\ee
When we associate $X$ with $\Phi$ we have
that
\be
\dot{X}=\dot{\Phi}
\ee
which to leading order yields after use of (\ref{e23})
\be
\dot{X}=\frac{ P}{M}\phi'_c
\ee
The equality holds exactly when
\be
\phi_c(x)=x\label{ap}
\ee
which is the expansion  of the soliton field about the kink position.
Thus when we use (\ref{dif}) we are taking (\ref{ap}) to be the solution of the
field
at arbitrary distances of the kink position, rather than the solution
(\ref{ex}).

Using the identity \cite{Ohta}
\be
< P'|\phi_c(x+X)| P>=\int dz \phi_c(z-x) e^{i \Delta  P' z}
e^{i \Delta E_{ P'} t}=
\phi_c(\Delta P') e^{i \Delta E_{ P'} t}
\ee
where $\Delta  P'= P- P'$ and $\Delta E_{ P'}=\frac{ P^2- P'^2}{2M}$,
we see that the equation (\ref{diff}) becomes for the forward term
\be
\Gamma_f(w)=\int dt\frac{\int d P' \Delta E_{ P'}^4 |\phi_c(\Delta P')|^2 e^{i
\Delta E_{ P'} t}}
{\int d P'' \Delta E_{ P''}^2 |\phi_c(\Delta P'')|^2 e^{i \Delta E_{ P''}
t}}\theta(t)e^{-i \om t}\label{gm1}
\ee
The backwards component  reads
\be
\Gamma_b(w)=\int dt \frac{\int d P' \Delta E_{ P'}^4 |\phi_c(\Delta P')|^2 e^{i
\Delta E_{ P'} t}}
{\int d P'' \Delta E_{ P''}^2 |\phi_c(\Delta P'')|^2 e^{i \Delta E_{ P''}
t}}\theta(-t)e^{-i \om t}\label{gm}
\ee
It follows from the construction of $\Gamma(w)$ that the equation (\ref{def})
for the soliton diffusion is mediated by phonons whose wavelength is $\Delta
P$
\cite{Gjs}.

We know that
soft acoustic phonons will not see the soliton profile because their wavelength
is much larger then the soliton extension.   On the other hand,  we see that
the
equations (\ref{gm}) and (\ref{gm1}) depend on the soliton profile.
Thus if we replace $\phi_c$ by $X$ we must
impose a cutoff on  the integrals over the momenta $ P'$ $ P''$.  The cutoff
should
be such that the smallest
wavelength is
much larger than the soliton extension.   However, if we do not make such a
replacement,
we may safely integrate over the phonon spectrum, optical phonons included.

\begin{figure}[hb]

\includegraphics{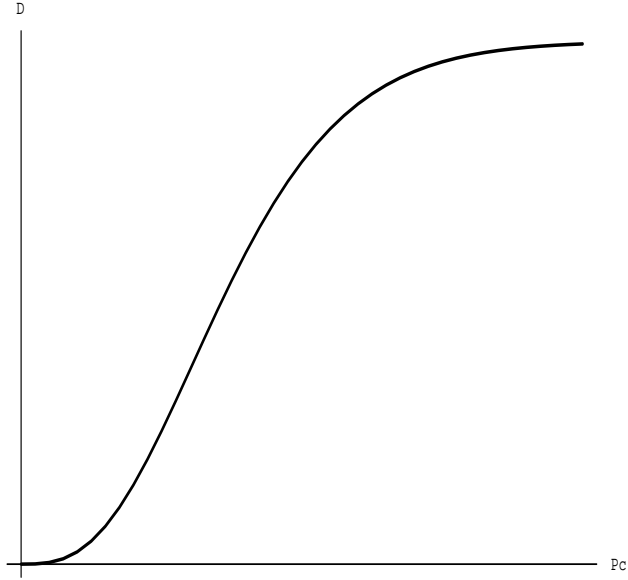}
\vspace{9.0cm}
\caption[]{Arbitrary Units}

\label{supergraph}
\end{figure}

To see the difference between using the classical solution (\ref{ex}) and using
the collective coordinate in the diffusion equation (\ref{def}) it is
convenient to
write (\ref{def}) as \cite{Wada}
\be
D_P(\om)=\int dt <P|\dot{X}(t)\dot{X}(0)|P>e^{-i\om t}.\label{de7}
\ee
If we replace the collective coordinate of the soliton with the soliton field
in
(\ref{de7}) we obtain
\be
D_P(\om)=\int dt <P|\dot{\Phi}(t)\dot{\Phi}(0)|P>e^{-i\om t}.\label{de1}
\ee
which after using the integral representation for the Heavy-side function and
integrating in time takes the form
\be
D_P(\om)= \frac{1}{i\pi} \int^{P_c}_{-P_c} d P' \frac{\Delta E^3_{
P'}|\phi'_c(\Delta P')|^2}{\Delta E^2_{ P'}-\om^2}.\label{def3}
\ee
We may then study how equation (\ref{def3}) behaves for different profiles
as a function of the cutoff.
First we note that since we are interested in the contribution of
optical phonons  and
as pointed out in \cite{Wad}, only the leading order term in the $\om$
expansion
represent the contribution from the optical phonons, we may discard terms
which depend on $\om$.  Unfortunately, we cannot get an analytical
expression for the profile (\ref{ex}).  However, assuming that $P=0$, we
may study how $D_0(0)\equiv D$ behaves as a function of the cutoff,
$P_c$, in the integral (\ref{def3}).  The behaviour is plotted in
Figure \ref{supergraph}.

We find that the diffusion equation saturates for high values of $P_c$ and
it is independent of $P_c$ for large $P_c$ allowing us to carry the integration
of $D$ to arbitrary large $P_c$.

Next we investigate the behaviour for the profile
$\Phi=X$.  It  will give us a taste of how sensible is the diffusion
equation (\ref{de1}) to the profile of the soliton.
In this case
the contribution from the optical phonons is obtained analyticaly
\be
D\sim \int^{P_c}_{-P_c} d P' sin^2(\oh L  P') \label{def55}
\ee
where $L>>1$ is the length of the box in which the soliton is confined.
Evaluation of the integral as a function of the cutoff shows that $D$
grows  like
\be
D\sim P_c
\ee
This dependence was expected and shows
that an improper   treatment of the collective coordinates leads to
a diffusion equation which is ill defined.
We also conclude that $D$ is quite sensible to
different profiles which lead to different contribution
from the optical phonons.  For the soliton profile we see that the contribution
of optical phonons saturates for large momenta.  However for  the profile $X$,
$D$ continues to increase linearly in $P_c$.
Define as cutoff the value of momenta for
 for which $D$ saturates when considering the correct soliton profile.
Then this must be the cutoff imposed  on
the profile $X$ to have a sensible theory.  Thus for acoustic phonons
the diffusion equation is not sensible to the profile.  However, for
optical phonons the diffusion equation  is very sensitive to the profile.

\section{Conclusions}

The existence of a constraint in the system we have considered
is a consequence of the broken
translational symmetry  of the vacuum, e.i. by our choice of
classical solution.
It has been shown that the operators defining the diffusion equation for the
soliton
must commute with the constraint.  Because the operators defining the diffusion
equation
in \cite{Wada} are not physical because they do not commute with the constraint
it
follows that this diffusion equation does not have a physical meaning.
In order to have only the
contributions from the physical sector of the theory we must replace the
operators used
in \cite{Wada} to define the soliton diffusion by physical ones.  The canonical
operator
which is physical and contains the information of the soliton position
is the soliton field.  We have also shown that using the  collective coordinate
as
the operator defining the soliton profile is not only unphysical but it is also
a
good approximation to the soliton profile if we consider the contribution of
the
acoustic  phonons only.  On the other hand, the contribution from the optical
phonons,
as we have shown, is very sensitive to the soliton profile.  One must therefore
use the exact profile of the soliton and not just the linear term in the
expansion
of the soliton profile when considering the contribution of the
optical phonons to  the diffusion of the soliton.

\vspace{2cm}

{\bf Acknowledgements}

It is a pleasure for me to thank Yu Lu for many interesting and
helpful conversations.

\pagebreak

\end{document}